\documentclass[runningheads]{llncs}
\usepackage{graphicx}
\usepackage{amsmath,amssymb} 
\usepackage{subcaption}
\usepackage{tabularx}
\usepackage{tikz}
\usepackage{cite}
\usepackage{marvosym}
\usepackage[utf8x]{inputenc}
\usepackage{algorithm}
\usepackage{algorithmic}
\usepackage{threeparttable}
\usepackage{tablefootnote}
\usepackage[normalem]{ulem}

\begin{document}
\title{Localizing Anatomical Landmarks in Ocular Images using Zoom-In Attentive Networks}

%
%
\author{Xiaofeng Lei\inst{1} \and
Shaohua Li\inst{1}\textsuperscript{\Letter} \and
Xinxing Xu\inst{1}\textsuperscript{\Letter} \and
Huazhu Fu\inst{1} \and
Yong Liu\inst{1} \and
Yih-Chung Tham\inst{2,3} \and
Yangqin Feng\inst{1} \and
Mingrui Tan\inst{1} \and
Yanyu Xu\inst{1} \and
Jocelyn Hui Lin Goh \inst{2} \and
Rick Siow Mong Goh \inst{1}\and
Ching-Yu Cheng \inst{2,3}
}
\authorrunning{X. Lei et al.}
\titlerunning{Zoom-In Attentive Networks for Localization}
%
\institute{Institute of High Performance Computing, A*STAR, Singapore \email{\{lei\_xiaofeng,li\_shaohua,xuxinx\}@ihpc.a-star.edu.sg}
\and
Singapore Eye Research Institute, Singapore National Eye Centre 
\and
Department of Ophthalmology, Yong Loo Lin School of Medicine, NUS, Singapore
}
\maketitle              
\begin{abstract}
Localizing anatomical landmarks are important tasks in medical image analysis. However, the landmarks to be localized often lack prominent visual features. Their locations are elusive and easily confused with the background, and thus precise localization highly depends on the context formed by their surrounding areas. 
In addition, the required precision is usually higher than segmentation and object detection tasks. Therefore, localization has its unique challenges different from segmentation or detection.
In this paper, we propose a zoom-in attentive network (ZIAN) for anatomical landmark localization in ocular images.
First, a coarse-to-fine, or ``zoom-in" strategy is utilized to learn the contextualized features in different scales.
Then, an attentive fusion module is adopted to aggregate multi-scale features, which consists of 1) a co-attention network with a multiple regions-of-interest (ROIs) scheme that learns complementary features from the multiple ROIs, 2) an attention-based fusion module which integrates the multi-ROIs features and non-ROI features.
We evaluated ZIAN on two open challenge tasks, i.e., the fovea localization in fundus images and scleral spur localization in AS-OCT images. Experiments show that ZIAN achieves promising performances and outperforms state-of-the-art localization methods. The source code and trained models of ZIAN are available at https://github.com/leixiaofeng-astar/OMIA9-ZIAN.

\end{abstract}
\section{Introduction}

\begin{figure}[!t]
\centering
\includegraphics[width=0.9\textwidth]{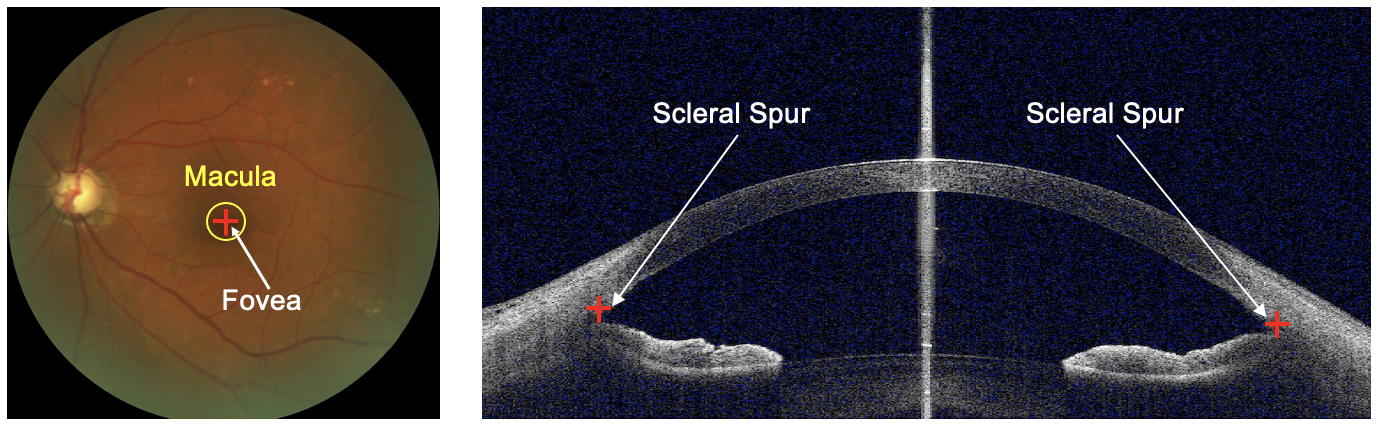}\centering
\caption{Two typical localization tasks in ocular images. Left: fovea location in fundus image. Right: Scleral Spur location in AS-OCT image.} \label{fig1}
\end{figure}

Localization of anatomical landmarks in medical images is one common task of medical image analysis. Precise localization plays an important role for some medical diagnosis. For example, the fovea is an important anatomical landmark on the posterior pole of the retina which is located in the center of a darker area of the eye~\cite{foveaclinal_info}. Fovea location is important in diagnosing eye diseases such as glaucoma, diabetic retinopathy and macular edema. Similarly, the Scleral Spur (SS) location is an important anatomical landmark in imaging the anterior chamber angle, as it is a reference point to identify open and narrow/closed angles based on Optical Coherence Tomography (OCT) images (Fig.~\ref{fig1}). 

Manually labeling these landmarks by medical experts is expensive and tedious. Developing automated approaches for landmark localization is desirable and has been studied for decades. The conventional computer vision methods mainly utilize template matching or mathematical morphology techniques to localize the anatomical landmark~\cite{math8050744,foveadetection_mia,foveadetection_icet,LI2021101971}. However, these methods are sensitive to the low contrast of the image and the results vary if the images come from a different source. With more robust performance, machine learning based approaches are predominantly used for automatic localization of anatomical landmarks~\cite{ml_in_medicine,zhao2019object,chen2020survey}. 

In general, there are three types of machine learning approaches for localization\cite{agereview}. 1) Localization is viewed as a value regression problem~\cite{Noothout_2020,DBLP:journals/corr/abs-2007-05355}, and the coordinates of the target location are directly predicted; 2) Localization is viewed as a binary segmentation problem that extends the single pixel label to a small region where the segmented mask center is used as the target position~\cite{vertebra_loz_miccai2021}; 3) Localization is viewed as heat-map regression task. First we generate a heat-map around the target position, and then employ regression, morphological or mathematical methods to estimate the target point~\cite{landmark_dl,co_attn_loz_miccai2021,app112110277,Liubjophthalmol-2021-319798,GU2Net}. Recently, the third heatmap-regression approach has outperformed the other 2 methods, and our method is also based on it.

Despite the huge progress in recent years, there are still challenges limiting the precision of these methods.
A common challenge is that input images may have highly varying scales. A second challenge is that anatomical landmarks often lack prominent visual features, and the localization highly depends on the context formed by their surrounding areas. 

In this paper, we propose ``Zoom-In Attentive Network" (ZIAN) to address the two challenges above, with ocular images as a case study. First, to be adaptive to various scales of input images, ZIAN adopts a zoom-in and a multi-scale ROI schemes; Second, to better incorporate surrounding areas as context for more precise localization, ZIAN adopts co-attention~\cite{Lu_2019_CVPR} and self-attention~\cite{segtran} mechanisms. 

In particular, different from the common ``zoom-in" strategy in~\cite{Liubjophthalmol-2021-319798,1e55b32f30894d9d96f9a9bde6d39f97} which predicts the final value more accurately based on the first approximation of the region in coarse stage, ZIAN utilizes a ``zoom-in" strategy, and a Regions-of-Interest (ROI) co-attention along with a self-attention mechanism that effectively fuses the multi-scales features in precise localization.
Specifically, in the zoom-in step, our model performs preliminary positioning of the target through a coarse network. As a result, multiple ROIs in different scales are cropped according to the preliminary position, which are used as the input to the fine network. In the attention step, a ROI co-attention~\cite{lu2017hierarchical,nguyen2018improved} module and a self-attention~\cite{transformer,analyze-multi-head,collab-multi-head,9193942} module work together to fuse the multi-ROI features. The ROI co-attention module fuses and complements the features of multi-ROIs. In addition, the self-attention module fuses the multi-ROI features with the output features from the coarse network for more accurate localization.
%
The main contributions of this paper are summarized as follows:
\begin{enumerate}
  \item Different from most existing localization frameworks, we present a ``Zoom-In Attentive Network" (ZIAN) that uses a coarse-to-fine zoom-in strategy, and a ROI co-attention/self-attention scheme in landmark localization.
  \item A novel attentive fusion module is proposed to adaptively fuse features from different ROIs, and then fuse the multi-scale ROI features with the coarse features, so that the model learns to combine features of multiple scales and multiple ROIs for better prediction.
\item We evaluated ZIAN on two common ocular image tasks, i.e.,  fovea localization in  fundus  images, and Scleral Spur (SS) localization in Anterior Segment Optical Coherence Tomography (AS-OCT) images. The effectiveness of the method is validated by comparing it with various state-of-the-art methods.
\end{enumerate}

\section{Method}
In this section, we provide details for our proposed Zoom-In Attentive Networks (ZIAN), which consists of two main components: the \emph{Zoom-in Module} and the \emph{Attentive Fusion Module} which includes the details of ROI co-attention and self-attention fusion module.
\subsection{Zoom-In Module}
\begin{figure}[!t]
\centering
\includegraphics[width=1\textwidth]{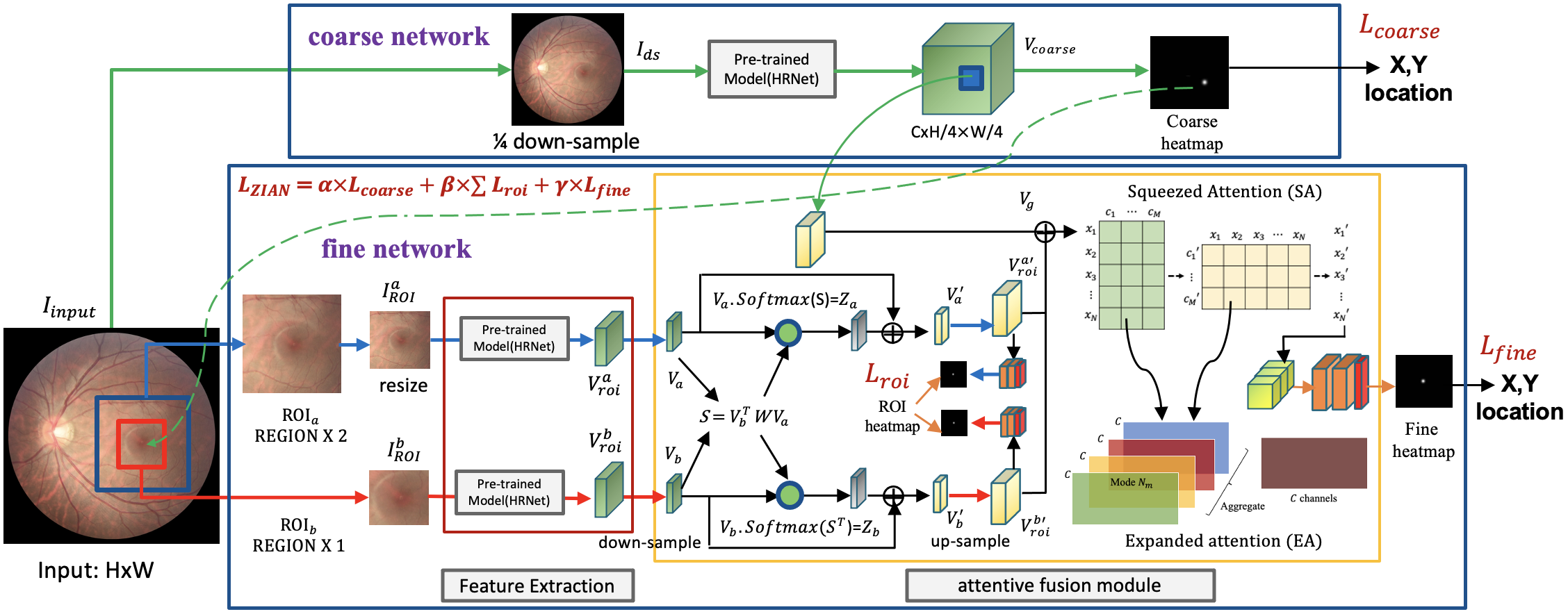}
\caption{The architecture of the proposed ZIAN, which comprises two main components: the \emph{Zoom-in Module} and the \emph{Attentive Fusion Module}. The input image is downsampled and fed into the coarse network to get the per-pixel coarse heat-map. The Multi-ROIs centered at the peak pixel on the coarse heat-map are cropped from the original input image, and fed into the fine network to generate features. Next, the multi-ROI features are refined by a co-attention module. Finally, the multi-ROI features are concatenated with the coarse-level features and transformed by self-attention module, and yield the fine heat-map.}\label{fig2}
\end{figure}

As shown in Fig.~\ref{fig2}, ZIAN has a coarse network and a fine network. The input image $I_{input}$ is down-sampled by 4$\times$ and fed into a pre-trained base network HRNet~\cite{HRNET} to get per-pixel heat-maps in the coarse network. The peak pixel is then located as the preliminary positioning of the target. Then, multiple scale ROIs centered at the preliminary location are cropped as the input of the fine network. The resized ROI images $I_{roi}^{a}$ and $I_{roi}^{b}$ are fed in parallel into the pre-trained model to build their feature representations individually. Next, multi-ROIs features $V_{roi}^{a}$ and $V_{roi}^{a}$ are processed through an attentive fusion module to get a fine-scale heat-map. The peak pixel in the fine-scale heat-map is located as the final coordinate of the target. 
We utilize HRNet~\cite{HRNET} as the pre-trained backbone in the figure. It can be replaced with any state-of-the-art backbone (U-Net~\cite{unet}, EfficientNet\cite{efficientnet}, YOLO~\cite{7780460}, RCNN~\cite{ren2016faster}, etc.).

\subsection{Attentive Fusion Module}
As shown at the bottom part in Fig.~\ref{fig2}, the fine network takes a pair of ROI images $I_{roi}^{a}$ and $I_{roi}^{b}$, which sequentially performs feature extraction and attentive fusion module which includes ROI co-attention and self-attention fusion. 
Two images $\rm{ROI}_{a}$ and $\rm{ROI}_{b}$ with different scales ($\times 1$ and $\times 2$, i.e. $256\times 256$ and $512\times 512$) are cropped on the input image $I_{input}$ and centered at the predicted peak pixel in coarse heat-map. The multi-ROIs features $V_{roi}^{a}$ and $V_{roi}^{a}$ extracted from pre-trained model are down-sampled and refined by the ROI co-attention module. 
Next, the refined multi-ROIs features $V_{a}^{'}$ and $V_{b}^{'}$ are up-sampled and concatenated with coarse-level features $V_{g}$ which we implement with "crop and resize" and "grid sample", then processed through a self-attention fusion module to get a fine-scale heat-map.
Our two-level attention mechanism in fine network guarantees the full integration of the features from different receptive fields, it maintains the independence and integrity of the individual CNN network for single ROI, which enables any CNN backbone to be implemented inside simultaneously.

\noindent \textbf{ROI Co-attention module}:
Since the input multi-ROIs images $\rm{ROI}_{a}$ and $\rm{ROI}_{b}$ are centred in the same preliminary positioning, and the feature extraction network is highly symmetric, we argue that multi-ROIs should have symmetric and complementary position representation which can guide each other to improve the discriminative ability of networks for landmark identification. We leverage the co-attention mechanism\cite{lu2017hierarchical,Lu_2019_CVPR} to mine the correlations between multi-ROIs features $V_{a}$ and $V_{b}$. We first compute the similarity matrix between ROI feature $V_{a}$ and $V_{b}$ ($\in\mathbb{R}^{W{\times}H{\times}C}$): ${S}=V_{b}^{T}WV_{a}\in\mathbb{R}^{WH{\times}WH}$,
where ${W\in\mathbb{R}^{C{\times}C}}$ is a weight matrix. Next, the attention summaries for the feature embedding can be computed as:
\begin{equation}
Z_{a}=V_{a}\cdot{Softmax(S)},\quad \text{and} \quad Z_{b}=V_{b}\cdot{Softmax(S^{T})}.
\end{equation}
We concatenate the co-attention representation $Z$ and the original ROI feature $V_{a}$ and $V_{b}$:  
    ${X_{a}=[Z_{a}, V_{a}], X_{b}=[Z_{b},V_{b}]}$  ($X_{a}, X_{b}\in \mathbb{R}^{W{\times}H{\times}2C}$).
Finally, $X_{a}$ and $X_{b}$ pass through a 3×3 convolution and batch norm followed by ReLU activation to get $V_{a}^{'}$ and $V_{b}^{'} ({\in \mathbb{R}^{W{\times}H{\times}C})}$ which keep the same 3D-tensor as $V_{a}$ and $V_{b}$. We apply downsampling and upsampling ($\times\frac{1}{4}$, $\times{4}$) before and after the ROI co-attention module to reduce memory footprint. Co-attention ROI feature $V_{roi}^{a'}$ and $V_{roi}^{b'}$ pass through a $1\times1$ convolution to output a landmark heat-map.

\noindent \textbf{Self-Attention Fusion module}:
The co-attention ROI feature $V_{roi}^{a'}$ and $V_{roi}^{b'}$ concatenated with "crop and resize" coarse-level features $V_{g}$, are fed to a self-attention fusion module. Self-attention module uses Squeezed Attention Block (SAB) and Expanded Attention Block (EAB) from Segmentation Transformers network\cite{segtran} so that our model can see the big picture in the features from the coarse network and fine details in the features from the fine network at the same time. 
SAB and EAB replace full self-attention and multi-head attention (MHA) in typical transformer to reduce noises and over-fitting in image tasks. SAB and EAB join forces to offer more capacity to model diverse data from coarse and fine networks.

The features $X_{out}$ after the Self-Attention Fusion module are followed by two 3×3 convolution and one convolution for the final heat-map. The peak value in the heat-map is located as the landmark position. Given the coordinates ($u_{0}$ , $v_{0}$) of landmark (Fovea or SS) point, the heat-map $G(u, v)$ as ground truth can be calculated as
    ${G(\boldsymbol{u}, \boldsymbol{v}) = \exp{(-\frac{(\boldsymbol{u}-\boldsymbol{u}_{0})^2 + (\boldsymbol{v}-\boldsymbol{v}_{0})^2}{2{\times}{\delta^2}})}}$, where $\delta$ is variance to control the heat-map radius, we use $\delta$=2 here.

The model is trained by minimizing the Mean Squared-Error (MSE) distance of the learned heat-map to a ground truth heat-map. Our ZIAN retains all loss functions of $L_{coarse}$ in the coarse network,  $L_{roi}$ and $L_{fine}$ in the fine network to improve their accuracy and combines them as
\begin{equation}
    \boldsymbol{L}_{ZIAN} = \alpha\times\boldsymbol{L}_{coarse}+\beta\times\sum\boldsymbol{L}_{roi},+\gamma\times\boldsymbol{L}_{fine},
\end{equation}
where $\boldsymbol{L}_{coarse}$, $\boldsymbol{L}_{roi}$ and $\boldsymbol{L}_{fine}$ are the MSE loss coming from coarse heat-map, ROI heat-map and fine heat-map in Fig.2 (ROI and fine ground truth heat-map are the same which are centered at the input ROI images). $\alpha$ , $\beta$ and $\gamma$ is the weight which is greater than or equal to 0 float, $\alpha$=1, $\beta$=0.25 and $\gamma$=1 here.
\section{Experiments and Results}
\subsection{Datasets and settings}
To validate the effectiveness of our model, we use the REFUGE dataset~\cite{refugedata} and AGE dataset~\cite{agereview}. The model is evaluated on two metrics: 1) Average Euclidean Distance (AVG L2) between the estimations and ground truth (unit is pixels, the lower the better) which is also the only evaluation criteria in two public challenges (REFUGE and AGE); 2) Successful detection rates (SDR, the higher the better)) with different thresholds (5 pixels, 10 pixels and 20 pixels).

REFUGE dataset\footnote{\url{https://refuge.grand-challenge.org}} consists of 1200 retinal fundus images ($1634\times 1634$) for fovea localization (400 train, 400 val, 400 test). We split 800 images (80$\%$:20$\%$) in training and evaluation.
AGE dataset\footnote{\url{https://age.grand-challenge.org}} consists of 4800 AS-OCT images for Scleral Spur (SS) localization (1600 train, 1600 val, 1600 test). We use those 1600 train images ($2130\times 998$) with publicly available ground truth (GT) in training and evaluation (80$\%$:20$\%$), 1600 val images out of other 3200 images(no GT released) for test.
The images are resized to $1064\times 1064$, and center cropped to $1024\times 1024$, then random cropped to a resolution of $896\times 896$, next downsampled to 1/4 of cropped image size, i.e. $224\times 224$ before being fed to pre-trained model in coarse network. In SS localization, we split each AS-OCT image into the left and right parts according to the centerline and locate the SS localization individually.

In SS localization task, GT of test dataset is not made public, all the results are obtained from online AGE Challenge Leaderboard for AVG L2 Distance.

\subsection{Experimental Setup}
ZIAN is implemented using PyTorch. All networks are trained using the Adam optimizer. We trained 140 epochs on the model with a learning rate of 0.0002, and weight decay of 0.1 after 90 epochs.
For data augmentation, we apply random horizontal flipping, drifting, scaling and rotation. The initial weights of the base networks are loaded from pre-trained models based on ImageNet, and the parameters of the other modules are randomly initialized.
We evaluate our ZIAN utilizing 2 state-of-the-art base networks: HRNet~\cite{HRNET} and U-Net~\cite{unet}. For each base network, we perform ablation studies to quantify the roles of different components namely base network, coarse-to-fine network with/without multi-ROIs scheme, self-attention or co-attention multi-ROIs scheme.
\subsection{Results and Discussion}
\begin{figure}[!t]
\includegraphics[width=1\textwidth]{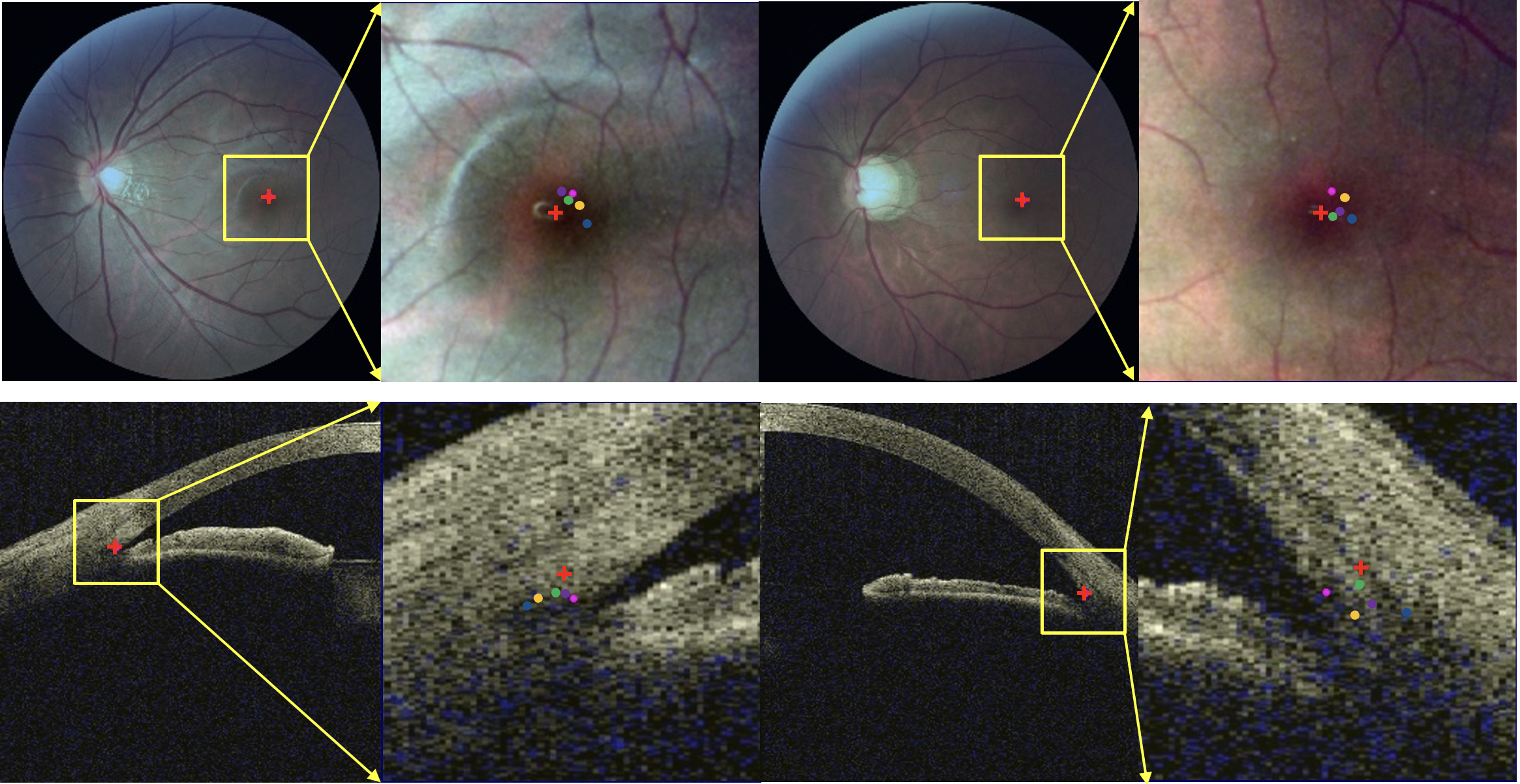}\centering 
\caption{The input and landmark locations in ZIAN coarse and fine network, Zoom-in localization results of ground truth and different methods. The red cross points {\color{red}+} are the ground truth labels, while the circle points are the learned landmarks, ${\color{blue}\bullet\color{black}}$, ${\color{yellow}\bullet\color{black}}$, ${\color{pink}\bullet\color{black}}$, ${\color{purple}\bullet\color{black}}$ and ${\color{green}\bullet\color{black}}$ are for "HRNet", "HRNet+1ROI", "HRNet+1ROI+SA", "HRNet+MR+SA" and "ZIAN with HRNet" respectively.} \label{fig3}
\end{figure}

In this part, we report the results of fovea and SS localization in the REFUGE and AGE test dataset using AVG L2 and SDR. The performances of different methods are reported in Table 1 and 2 with some results in the REFUGE and AGE challenge leaderboard. Some output examples from the coarse and fine network are as Fig.~\ref{fig3}.

In the tables, HRNet, U-Net, EffcientUNet and GU2Net refers to the method and the CNN backbone we utilize. ``1ROI" or ``MR" means that we adopt a Coarse-to-Fine strategy with 1 ROI or multi-ROIs in fine network, "SA" means only the self-attention part is employed in the attentive fusion module. ``ZIAN with HRNet/U-Net" is our proposed architecture using a coarse-to-fine strategy, a ROI co-attention along with a self-attention fusion mechanism. ``SDSA and VRT team" and ``Dream Sun and MIPAV team" are the top 2 in the REFUGE final rankings and AGE semifinal rankings, respectively, their results are presented for fair comparison. The corresponding methods of these teams are described in the REFUGE and AGE overview papers~\cite{refugedata,agereview} and website.

From the results, we could have several observations:

\begin{table}[!t]
\centering
\caption{Performances of different methods on REFUGE dataset. The best results are in \textbf{bold} and the second best results are \underline{underlined}, - represents that no experimental results can be found in REFUGE challenge.}\label{Fovea}
\begin{threeparttable}
\begin{tabular}{|l|c|c|c|c|}
\hline
Method & AVG L2 & SDR 5px(\%) & SDR 10px(\%) & SDR 20px(\%)\\
\hline
REFUGE SDSA team\tnote{+}&  \emph{34.7} & - & - & -\\
REFUGE VRT team\tnote{+}&  \emph{37.1} & - & - & -\\
U-Net\cite{unet} &  18.99 & 23.5 & 52.25 & 69.50\\
HRNet\cite{HRNET} &  14.45 & 30.0 & 70.50 & 88.75\\
EffcientUNet-B0\cite{efficientnet}\tnote{*} &  18.62 & 24.5 & 59.00 & 72.50\\
EffcientUNet-B5\cite{efficientnet}\tnote{*} &  26.29 & 15.5 & 40.75 & 55.75\\
GU2Net\cite{GU2Net} &  24.91 & 24.0 & 47.5 & 60.75\\
\hline
U-Net+1ROI &  16.56 & 29.75 & 61.0 & 78.75\\
U-Net+1ROI+SA &  15.05 & 30.25 & 62.5 & 78.0\\
U-Net+ MR +SA &  14.00 & 25.25 & 61.25 & 80.5\\
ZIAN W/U-Net (ours) &  13.24 & 32.0 & 67.75 & 82.0\\
\hline
HRNet+1ROI &  13.93& 28.75 & 67.25 & 84.25\\
HRNet+1ROI+SA &  9.51& \underline{41.25} & \textbf{79.25} & 90.0 \\
HRNet+ MR +SA &  \underline{9.42} & 38.0 & 76.25 & \underline{91.75} \\
ZIAN W/HRNet (ours) &  \textbf{9.07}& \textbf{44.25} & \underline{78.5} & \textbf{92.5}\\
\hline
\end{tabular}
\begin{tablenotes}
    \footnotesize
    \item[+] Top 2 teams in the REFUGE 1 final leaderboard. The model is trained on 400 train images instead of 800 train+val images like other methods.
    \item[*] \url{https://github.com/zhoudaxia233/EfficientUnet-PyTorch} 
\end{tablenotes}
\end{threeparttable}
\end{table}

\begin{table}[!t]
\centering
\caption{Performances of different methods on AGE dataset}\label{SS}
\begin{threeparttable}
\begin{tabular}{|l|c|l|c|c|}
\hline
Method &    AVG L2 & Method & AVG L2 & Params (M)\tnote{+}\\
\hline
AGE Dream Sun team\tnote{*} &  \emph{12.897} & & &\\
AGE MIPAV team\tnote{*} &  13.761 & HRNet\cite{HRNET} &  15.032 & 28.544 \\
EffcientUNet-B0\cite{efficientnet}\tnote{**} &  17.657 & HRNet+1ROI &  14.783 & 57.102 \\
EffcientUNet-B5\cite{efficientnet}\tnote{**}&  15.594 & HRNet+1ROI+SA &  14.264 & 57.125 \\
U-Net\cite{unet} &  21.257 & HRNet+ MR +SA &  13.891 & 85.705 \\
GU2Net\cite{GU2Net} &  23.024 & ZIAN W/HRNet (ours)&  \textbf{13.638} & 85.710 \\
\hline
\end{tabular}
\begin{tablenotes}
    \footnotesize
    \item[*] Top 2 teams in the AGE semi-final leaderboard, Dream Sun team utilizes Ensemble models with EffcientNet B2, B3, B5, and B6\cite{agereview}. 
    \item[**] \url{https://github.com/zhoudaxia233/EfficientUnet-PyTorch} 
    \item[+] Number of parameters in ablation study of ZIAN on a $224\times 224$ input image.
\end{tablenotes}
\end{threeparttable}
\end{table}

\textbf{Role of Coarse-to-Fine Strategy (C2F)}: We concur that the C2F strategy has been commonly used in localization and segmentation tasks to narrow down the ROI areas~\cite{agereview,9193942}. However, as indicated by our experiments, the C2F design is just a minor contributor for improvements compared with the other components (which are our main contribution) as described below.
 
\textbf{Role of Multi-ROIs (MR)}: The localization accuracy is sensitive to the choices of the cropped ROI sizes which confine the context. MR avoids manually tuning the choice of ROI sizes which is more robust with better coverage and applicable to a wide range of tasks, relieving people from ad-hoc tuning. With a MR scheme, the model can choose from multiple contexts and learn to construct more predictive features. Multi-ROIs can achieve superior performance as compared to that of 1 ROI as demonstrated in Tables 1 and 2.

\textbf{Role of Self-Attention Fusion module (SA)}: SA is to learn to fuse the multi-scales features, so that the model learns to combine features for better prediction. With SA as shown in Tables 1 and 2 with HRNet, it significantly reduces the L2 distance further from 13.93 to 9.51, and from 14.786 to 14.264, in Fovea and SS tasks, respectively. 
In order to further investigate the advantages of multi-ROIs with SA, we evaluate the impact of SA without MR scheme, i.e. HRNet+1ROI+SA vs HRNet+MR+SA. This model achieved L2 distances of 9.42 and 13.891, compared to 9.52 and 14.264 in Fovea and SS tasks respectively. It indicates that self-attention works well under a single ROI and multi-ROIs. 

\textbf{Role of ROI Co-Attention (RCA)}: RCA mines the underlying correlations between Multi-ROIs features and selectively focuses on landmark regions. With RCA as shown in Tables 1 and 2, it slightly reduces the L2 distance from 9.42 to 9.07, and from 13.891 to 13.638, in Fovea and SS tasks, respectively.

\textbf{Computational Efficiency}: After incorporating the MR+SA+RCA in fovea localization, the GPU RAM usage increased from 2GB to 11.9GB with HRNet backbone, the training speed decreased from 12 images/s to 4 images/s, and the test speed decreased from 50 images/s to 7.4 images/s on one workstation with NVIDIA RTX3090 graphics card. Table 2 presents the number of parameters of our ZIAN method. As the top priority of medical applications is accuracy, we think the computational overhead of the ZIAN is still feasible and manageable.

\section{Conclusions}
In this paper, we propose a Zoom-In Attentive Network (ZIAN) for landmark localization tasks. ZIAN consists of a  coarse-to-fine ``zoom-in" module and an attentive fusion module. In the attentive fusion module, a ROI co-attention along with a self-attention fusion combine and fuse the multi-scale multi-ROI features. We performed extensive experiments and ablation studies on two public ocular image datasets. The results demonstrate that ZIAN has advantages over commonly used baselines. In the future work, we would like to extend ZIAN to make it robust against domain distribution shifts of the input images.

\subsection*{Acknowledgements. 
{\normalfont This work was supported by the Agency for Science, Technology and Research (A*STAR) under its AME Programmatic Funds (Grant Number : A20H4b0141), its Career Development Fund (Grant No. C210112016), and its RIE2020 Health and Biomedical Sciences (HBMS) Industry Alignment Fund Pre-Positioning (IAF-PP, Grant Number : H20c6a0031). Xinxing Xu and Shaohua Li are the corresponding authors.}}

\bibliographystyle{ieeetr}
\bibliography{mybib}
\end{document}